\documentclass[12pt]{article}
\usepackage{amsfonts}
\usepackage{amsmath}
\usepackage{amssymb}
\usepackage{graphicx}
\usepackage{amssymb}
\usepackage{graphics}
\usepackage{epsfig}

\numberwithin{equation}{section}
\providecommand{\openone}{\leavevmode\hbox{\small1\kern-3.8pt\normalsize1}}
\newcommand{\Tr}{\mbox{Tr}}
\newcommand{\Det}{\mbox{Det}}
\newcommand{\rank}{\mbox{rank}}
\newtheorem{thm}{Theorem}

\newtheorem{defn}[thm]{Definition}

\begin{document}

\title{One-mode Bosonic Gaussian channels: a full weak-degradability classification}

\author{F. Caruso, V. Giovannetti  \\
{\small NEST CNR-INFM \& Scuola Normale Superiore,} \\
{\small Piazza dei Cavalieri 7, I-56126 Pisa, Italy}
 \\ \\
A. S. Holevo\thanks{The Leverhulme Visiting Professor at
Centre for Quantum Computation, Department of Applied Mathematics and Theoretical Physics, Cambridge University.} \\
{\small Steklov Mathematical Institute,}
\\ {\small Gubkina 8,
119991 Moscow, Russia}}
\maketitle

\begin{abstract}
A complete degradability analysis of one-mode Gaussian Bosonic
channels is presented. We show that apart from the class of
channels which are unitarily equivalent to the channels with
additive classical noise, these  maps can be characterized in
terms of weak-  and/or anti-degradability. Furthermore a new set of
channels which have null quantum capacity is identified. This is
done by exploiting the composition rules of one-mode Gaussian maps
and the fact that anti-degradable channels can not be used to
transfer quantum information.
\end{abstract}

\maketitle

Within the context of quantum information theory~\cite{SHOR}
Bosonic Gaussian channels~\cite{HPPI,HW,REV} play  a fundamental
role. They include all the physical transformations which preserve
``Gaussian character'' of the transmitted signals and can be seen
are the quantum counterpart of the Gaussian channels in the
classical information theory~\cite{GALLAGER}. Bosonic Gaussian
channels describe most of the noise sources which  are routinely
encountered in optics, including those responsible for the
attenuation and/or the amplification of signals along optical
fibers. Moreover, due to their relatively simple structure, these
channels provide an ideal theoretical playground for the study of
continuous variable~\cite{CONVAR} quantum communication protocols.

Not surprisingly in the recent years an impressive effort has been
put forward to characterize the properties of Bosonic Gaussian
channels. Most of the efforts focused on the evaluation of the
optimal transmission rates of these maps under the constraint on
the input average energy both in the multi-mode scenario (where
the channel acts on a collection of many input Bosonic mode) and
in the one-mode scenario (where, instead, it operates on a single
input Bosonic mode). In few cases~\cite{CAVES,HSH,LOSS,BROAD} the
exact values of the communication
capacities~\cite{HSW,HOLEVOSHIROKOV,SETH} of the channels have
been computed. In the general case however only certain bounds are
available (see~\cite{HW,BROAD,BENIEEE,HALL,ENTROPY}). Finally
various additivity issues has been analyzed in
Refs.~\cite{GL,SEW}.

Recently the notions of  anti-degradability and weak-degradability
were proposed as an useful tool for studying the quantum capacity
properties of one-mode Gaussian channels~\cite{CG}. This suggested
the possibility of classifying these maps in terms of a simple
canonical form which was achieved in Ref.~\cite{HOLEVOREP}.
Moreover, proceeding along similar lines,  the exact solution of
the quantum capacity of an important subset of those channels was
obtained in Ref.~\cite{WOLF}.

In this paper we provide a complete degradability classification
of one-mode Gaussian channels and exhibit a new set of channels
which have null quantum capacity extending a previous result in
Ref.~\cite{HW}.

The definition of weak-  and anti-degradability of a quantum
channel is similar to the definition of degradability introduced
by Devetak and Shor in Ref.~\cite{DEVSHOR}. It is based on
replacing the Stinespring dilation~\cite{STINE} of the channel
with a representation where the ancillary system (environment) is
not necessarily in a {\em pure} state~\cite{HPPI,LINDBLAD}. This
yields a generalization of the notion of {\em complementary}
channel from Ref.~\cite{DEVSHOR,CONJ0,CONJ} which is named {\em
weakly complementary}~\cite{CG}. In this context weakly degradable
are those channels where the modified state of the ancillary
system -- described by the action of the weakly complementary
channel -- can be recovered from the output state of the channel
through the action of a third channel. Vice-versa anti-degradable
channels obey the opposite rule (i.e. the output state of the
channel can be obtained from the modified state of the ancilla
through the action of another
 suitable channel).
Exploiting the canonical form~\cite{HOLEVOREP} one can show that,
apart from the class $B_2$  consisting of the maps which are
unitarily equivalent to the channels with additive classical
Gaussian noise~\cite{HW}, all one-mode Bosonic Gaussian channels
are either weakly degradable or anti-degradable. As discussed in
Ref.~\cite{CG} the anti-degradability property allows one to
simplify the analysis of the quantum capacity~\cite{SETH} of these
channels. Indeed those maps which are anti-degradable can be shown
to have null quantum capacity. On the other hand, those channels
which are weakly degradable with pure ancillas (i.e. those which
are degradable in the sense of Ref.~\cite{DEVSHOR}) have quantum
capacity which can be expressed in terms of a single-letter
expression. Here we will focus mostly on the anti-degradability
property, and, additionally, we will show that by exploiting the
composition rules of one-mode Bosonic Gaussian channels, one can
extend the set the maps with null quantum capacity well beyond the
set of anti-degradable maps.

The paper is organized as follows. In Sec.~\ref{s:uno} we
introduce the notion of weakly complementarity and weak-degradability 
in a rather general context. In Sec.~\ref{s:gaus} we
give a detailed description of the canonical decomposition of
one-mode Bosonic Gaussian channels. In Sec.~\ref{s:DEGRADABILE} we
discuss the weak-degradability properties of one-mode channels.
Finally, in Sec~\ref{sec:null} we
 determine the new set  of
channels with null quantum capacity.

\section{Weakly complementary and weakly degradable channels}~\label{s:uno}

In quantum mechanics, quantum channels describe evolution of an
open system $A$ interacting with external degrees of freedom. In
the Sch\"{o}dinger picture these transformations are described by
completely positive trace preserving (CPT) linear maps $\Phi$
acting on the set ${\cal D}({\mathcal H_a})$ of the density
matrices $\rho_a$ of the system. It is a well known (see e.g.
\cite{HPPI}, \cite{LINDBLAD}) that $\Phi$ can be described by
 a unitary coupling between the system $A$ in input state $\rho_a$  with an external
ancillary system $B$ (describing the {\em environment}) prepared
in some fixed {\em pure} state. This follows from Stinespring
dilation~\cite{STINE} of the map which is unique up to a partial
isometry. More generally, one can describe $\Phi$ as a coupling
with environment prepared in some {\em mixed} state $\rho_b$, i.e.
 \begin{eqnarray}
\Phi(\rho_a) = \mbox{Tr}_b[ U_{ab} (\rho_a\otimes \rho_b)
U_{ab}^\dag] \;, \label{HGCuno}
\end{eqnarray}
where $\Tr_b [ ... ]$ is the partial trace over the environment
$B$, $U_{ab}$ is a unitary operator in the composite Hilbert space
${\cal H}_a\otimes {\cal H}_b$. We call Eq.~(\ref{HGCuno}) a
``physical representation'' of $\Phi$ to distinguish it from the
Stinespring dilation, and to stress its connection with the
physical picture of the noisy evolution represented by $\Phi$. Any
Stinespring dilation gives rise to a physical representation.
Moreover from any physical representation~(\ref{HGCuno}) one can
construct a Stinespring dilation by purifying $\rho_b$ with an
external ancillary system $C$, and by replacing $U_{ab}$ with the
unitary coupling $U_{abc} = U_{ab}\otimes \openone_{c}$.

Equation~(\ref{HGCuno})  motivates the following~\cite{CG}
\begin{defn}
For any physical representation {\em (\ref{HGCuno})} of  the
quantum channel $\Phi$ we define its {\em weakly complementary} as
the map ${\tilde{\Phi}}:{\cal D}({\cal H}_a) \rightarrow {\cal
D}({\cal H}_b)$  which takes the input state $\rho_a$ into the
 state of the
environment $B$ after the interaction with $A$, i.e.
\begin{eqnarray}
\tilde{\Phi}(\rho_a) = \mbox{\em Tr}_a[ U_{ab} (\rho_a\otimes
\rho_b) U_{ab}^\dag] \;. \label{duedue}
\end{eqnarray}
\end{defn}
The transformation~(\ref{duedue}) is CPT, and it describes a
quantum channel connecting systems $A$ and $B$. It is a
generalization of the {\em complementary (conjugate) channel}
$\Phi_{\text{com}}$ defined in Ref.~\cite{DEVSHOR,CONJ0,CONJ}. In
particular, if Eq.~(\ref{HGCuno}) arises from a Stinespring
dilation (i.e. if $\rho_b$ of Eq.~(\ref{duedue}) is pure) the map
$\tilde{\Phi}$ coincides with $\Phi_{\text{com}}$. Hence the
latter is a particular instance of a  weakly complementary channel
of $\Phi$. On the other hand, by using the above purification
procedure, we can always represent a weakly complementary map as a
composition
\begin{equation}\label{comp}
\tilde{\Phi}=T\circ\Phi_{\text{com}},
\end{equation}
where $T$ is the partial trace over the purifying system (here
{\em `` $\; \circ$ ''} denotes the composition of channels).
 As we will see, the
properties of weakly complementary and complementary maps in
general differ.

\begin{defn}
Let  $\Phi, \tilde{\Phi}$ be a pair of mutually
weakly-complementary channels such that
\begin{eqnarray}
({\Psi}\circ {\Phi})(\rho_a) = \tilde{\Phi}(\rho_a) \;,\label{deg}
\end{eqnarray}
for some channel $ \Psi : {\cal D}({\cal H}_a) \rightarrow {\cal
D}({\cal H}_b)$ and all density matrix $\rho_a \in {\cal D}({\cal
H}_a)$. Then $\Phi$ is called  {\em weakly-degradable}  while
$\tilde{\Phi}$ -- {\em anti-degradable} (cf.~\cite{CG}).
\end{defn}
Similarly if
\begin{eqnarray}
(\overline{\Psi}\circ \tilde{\Phi})(\rho_a) = {\Phi}(\rho_a) \;,\label{antideg}
\end{eqnarray}
for  some channel $ \overline{\Psi} : {\cal D}({\cal H}_b)
\rightarrow {\cal D}({\cal H}_a)$ and all density matrix $\rho_a \in
{\cal D}({\cal H}_a)$, then  $\Phi$ is {\em anti-degradable} while
$\tilde{\Phi}$ is {\em weakly-degradable}.

In Ref.~\cite{DEVSHOR} the channel $\Phi$ is called {\em
degradable} if in Eq.~(\ref{deg}) we replace $\tilde{\Phi}$ with a
complementary map $\Phi_{\text{com}}$ of $\Phi$. Clearly any
degradable channel~\cite{DEVSHOR} is weakly degradable but the
opposite is not necessarily true. Notice, however, that due to
Eq.~(\ref{comp}), in the definition of anti-degradable channel we
can always replace weakly complementary with complementary (for
this reason there is no point in introducing the notion of weakly
anti-degradable channel). This allows us to verify that if $\Phi$
is anti-degradable~(\ref{antideg}) then its complementary channel
$\Phi_{\text{com}}$ is degradable~\cite{DEVSHOR}
 and vice-versa.
It is also worth pointing out that channels which are unitarily
equivalent to a channel $\Phi$ which is weakly degradable
(anti-degradable) are also weakly degradable (anti-degradable).

Finally an important property of anti-degradable channels is the
fact that their quantum capacity~\cite{SETH} is null. As discussed
in~\cite{CG} this is a consequence of the no-cloning theorem~\cite{NOCLONING}
(more precisely, of the impossibility of cloning with arbitrary high
fidelity~\cite{NOCLONING2}).

It is useful also to reformulate our definitions in the Heisenberg
picture. Here the states of the system are kept fixed and the
transformation induced on the system by the channel is described
by means of a linear map $\Phi_H$ acting on the algebra ${\cal
B}({{\cal H}_a})$ of all bounded operators of $A$ so that
\begin{eqnarray}
\Tr_a [ \Phi(\rho_a) \; \Theta_a  ] = \Tr_a [ \rho_a \;  \Phi_H
(\Theta_a)] \label{identity}\;,
\end{eqnarray}
for all $\rho_a\in {\cal D}({\cal H}_a)$ and for all
 $\Theta_a \in {\cal B}({\cal H}_a)$.
From this it follows that the Heisenberg picture counterpart of
the physical representation~(\ref{HGCuno})  is given by the unital channel
\begin{eqnarray}
\Phi_H(\Theta_a) &=&  \mbox{Tr}_b \big[  \; U_{ab}^\dag\;
(\Theta_a \otimes \openone_b) \; U_{ab} \; (\openone_a \otimes \rho_b
) \; \big]\;. \label{physicalHEIS}
\end{eqnarray}
Similarly, from~(\ref{duedue}) it follows that in the Heisenberg
picture the weakly complementary of the channel is described by
the completely positive unital map
\begin{eqnarray}
\tilde{\Phi}_H(\Theta_b) = \mbox{Tr}_b \; \big[ \; U_{ab}^\dag
(\openone_a \otimes \Theta_b) \; U_{ab} \; \big( \openone_a
\otimes \rho_b) \; \big] \label{wconjHeis} \;,
\end{eqnarray}
which takes bounded operators in ${\cal H}_b$ into bounded
operators in ${\cal H}_a$.

Within this framework the weak-degradability property~(\ref{deg})
of the channel $\Phi_H$ requires the existence of a channel
$\Psi_H$ taking bounded operators of ${\cal H}_b$ into bounded
operators of ${\cal H}_a$, such that
\begin{eqnarray}
(\Phi_H \circ \Psi_H)(\Theta_b) = \tilde{\Phi}_H(\Theta_b)
\;,\label{wwwweak}
\end{eqnarray}
for all $\Theta_b \in {\cal B}({\cal H}_b)$. Similarly we say that a
quantum channel $\Phi_H$ is anti-degradable, if there exists a
channel $\overline{\Psi}_H$ from ${\cal B}({\cal H}_a)$ to ${\cal
B}({\cal H}_b)$, such that
\begin{eqnarray}
(\tilde{\Phi}_H \circ \overline{\Psi}_H) (\Theta_a)
={\Phi}_H(\Theta_a) \;, \label{wwwanti}
\end{eqnarray}
for all $\Theta_a \in {\cal B}({\cal H}_a)$.


\section{One-mode Bosonic Gaussian channels}\label{s:gaus}

Gaussian channels arise from linear dynamics of open Bosonic
system interacting with Gaussian environment via quadratic
Hamiltonians. Loosely speaking, they can be characterized as CPT
maps that transform Gaussian states into Gaussian
states~\cite{HW,REV,REV1}. Here we focus on one-mode Bosonic
Gaussian channels which act on the density matrices of single
Bosonic mode $A$. A classification of such maps obtained recently
in the paper~\cite{HOLEVOREP} allows us to simplify the analysis
of the weak-degradability property. In the following we start by
reviewing the result of Ref.~\cite{HOLEVOREP}, clarifying the
connection with the analysis of Ref.~\cite{CG} (cf. also
Ref.~\cite{SEW}). Then we pass to the weak-degradability analysis
of these channels, showing that with some important exception,
they are either weakly degradable or anti-degradable.

\subsection{General properties}

Consider a single Bosonic mode characterized by canonical
observables $Q_a, P_a$ obeying the canonical commutation relation
$[Q_a,P_a]=i$. A consistent description of the system can be given
in terms of the unitary Weyl operators $V_a(z)= \exp \,[ i(Q_a,
P_a)\cdot z ]$, with
 $z= (x,y)^T$  being a column vector of $R^2$. In this
framework the canonical commutation relation is  written as
\begin{equation*}
V_a(z)\;V_a(z^{\prime })=\exp [\frac{i}{2}\Delta (z,z^{\prime
})]\; V_a(z+z^{\prime })\;, \label{weyl}
\end{equation*}
where $\Delta (z,z^{\prime })$
is  the symplectic form
\begin{equation}
\Delta (z,z^{\prime })= -i\;  z^T \cdot \sigma_2 \cdot z^\prime
=x^{\prime }y-xy^{\prime }\;,  \label{sympl-form}
\end{equation}
with $\sigma_2$ being the second Pauli matrix. Moreover the
density operators $\rho_a$ of the system can be expressed in terms
of an integral over $z$ of the $V_a(z)$'s, i.e.
\begin{eqnarray}
\rho_a = \int \frac{d^2 z}{2\pi} \; \phi(\rho_a; z) \; V_a(-z) \;,
\label{decomposition}
\end{eqnarray}
with
\begin{eqnarray}
\phi( \rho_a; z) = \Tr_a [ \rho_a \; V_a(z) ]\;, \label{characte}
\end{eqnarray}
being the characteristic function of $\rho_a$~\footnote{ In effect
an analogous decomposition~(\ref{decomposition}) holds also for
all trace class operators of $A$~\cite{HOLEVOBOOK}.}. Consequently
a complete description of a quantum channel on $A$ is obtained by
specifying its action on the operators $V_a(z)$, or, equivalently,
by specifying how to construct the characteristic function $\phi(
\Phi(\rho_a); z)$ of the evolved states. In the case of Gaussian
channels $\Phi$ this is done by assigning a mapping of the Weyl
operators
\begin{equation}
\Phi_H(V_a(z))=V_a(K\cdot z )\; \exp[ -\frac{1}{2}\; z^T \cdot
\alpha
 \cdot z  + i \; m^T
\cdot  z] \label{linbos}\;,
\end{equation}
in the Heisenberg picture, or the transformation of the
characteristic functions
\begin{eqnarray}
\phi(\Phi(\rho_a);z) =\phi(\rho_a; K \cdot z)\; \exp[
-\frac{1}{2}\; z^T \cdot \alpha \cdot z + i \; m^T \cdot z ],
\label{linbos1}
 \end{eqnarray}
in the Schr\"odinger picture.
Here  $m$ is a vector, while $K$ and $\alpha$ are real matrices
(the latter being symmetric and positive).
Equation~(\ref{linbos1}) guarantees that any input Gaussian
characteristic function will remain Gaussian under the action of
the map. A useful property of Gaussian channels is the fact that
the composition of two of them (say $\Phi^{\prime}$ and
$\Phi^{\prime\prime}$) is still a Gaussian channel. Indeed one can
easily verify that the composite map $\Phi^{\prime\prime}\circ
\Phi^{\prime}$ is of the form~(\ref{linbos1}) with $m$, $K$ and
$\alpha$ given by
\begin{eqnarray}
m &=& (K^{\prime\prime})^T \cdot m^{\prime} + m^{\prime\prime} \nonumber \\
K &=& K^{\prime} \; K^{\prime\prime} \label{composition} \\
\alpha &=& (K^{\prime\prime})^T \; \alpha^\prime \;
K^{\prime\prime} + \alpha^{\prime\prime}\;. \nonumber
\end{eqnarray}
Here $m^\prime$, $K^\prime$, and $\alpha^\prime$ belongs to
$\Phi^\prime$ while $m^{\prime\prime}$, $K^{\prime\prime}$, and
$\alpha^{\prime\prime}$ belongs to $\Phi^{\prime\prime}$.

Not all possible choices of $K$, $\alpha$ correspond to
transformations $\Phi$ which are completely positive. A necessary
and sufficient condition for this last property (adapted to the
case of one mode) is provided by the nonnegative definiteness of
the following $2\times 2$ Hermitian matrix~\cite{HW,HOLEVOREP}
\begin{equation}\label{positive}
2 \; \alpha -  \sigma_2 +K^T \; \sigma_2 \; K \;.
\end{equation}
This  matrix reduces to $2 \alpha + (\Det[K] -1) \; \sigma_2$ and
its nonnegative definiteness to
the inequality
\begin{eqnarray}
\Det[\alpha] \geqslant \left(\frac{\Det[K]-1}{2}\right)^2 \;. \label{inequality}
\end{eqnarray}
Within the limit imposed by Eq.~(\ref{inequality})
we can use Eq.~(\ref{linbos1}) to describe the whole
set of the one-mode Gaussian channels.

\subsection{Channels with single-mode physical
representation}\label{s:single} An important subset of one-mode
Gaussian  channels is given by the maps $\Phi$ which possess a
physical representation~(\ref{HGCuno}) with $\rho_b$ being a
Gaussian state of a {\em single} external Bosonic mode $B$ and
with $U_{ab}$ being a canonical transformation of $Q_a$, $P_a$,
$Q_b$ and $P_b$ (the latter being the canonical observables of the
mode $B$). In particular let $\rho_b$ be a thermal state of
average photon number $N$, i.e.
\begin{eqnarray}
\phi(\rho_b;z) = \Tr_b [ \rho_b \;V_b(z)  ] = \exp[ - (N+1/2)
|z|^2/2] \; \label{sigmab},
\end{eqnarray}
and let $U_{ab}$ be such that
\begin{eqnarray}
U_{ab}^\dag \;(Q_a , P_a, Q_b, P_b) \; U_{ab} = (Q_a, P_a, Q_b,
P_b) \cdot M \;, \label{couplingN}
\end{eqnarray}
with $M$ being a $4\times 4$ symplectic matrix of  block  form
\begin{eqnarray}
M \equiv \left( \begin{array}{ccc}
m_{11}&|& m_{21}  \\ \hline
m_{12}&|& m_{22}
\end{array} \right)\;. \label{matricem}
\end{eqnarray}
This yields the following evolution for the characteristic
function $\phi(\rho_a;z)$,
\begin{eqnarray}
\phi(\Phi(\rho_a); z) &=& \Tr_a[ \Phi(\rho_a) \; V_a(z) ] = \Tr_a [ \rho_a \;\Phi_H(V_a(z))]  \nonumber \\
&=& \Tr_{ab} \left[ U_{ab}^\dag \; ( V_a(z) \otimes \openone )
U_{ab} \; ( \rho_a \otimes \rho_b) \right]
\nonumber \\
&=& \Tr_{ab} \left[\big( V_a(m_{11} \cdot z) \otimes
V_b(m_{12}\cdot z) \big) \; ( \rho_a \otimes \rho_b) \right]
\nonumber \\
&=&  \phi(\rho_a; m_{11} \cdot z) \; \exp[ - (N+1/2) | m_{12}\cdot
z|^2 /2]\;, \label{output}
\end{eqnarray}
which is of the form~(\ref{linbos1}) by choosing $m=0$, $K=m_{11}$
and $\alpha= (N+1/2) \; m_{12}^T \cdot m_{12}$. It is worth
stressing that in the case of Eq.~(\ref{output}) the
inequality~(\ref{inequality}) is guaranteed by the symplectic
nature of the matrix $M$, i.e. by the fact that
Eq.~(\ref{couplingN}) preserves the commutation relations among
the canonical operators. Indeed we have
\begin{eqnarray}
\Det[\alpha] &=& (N+1/2)^2 \; \Det[m_{12}]^2 = (N+1/2)^2 \; (\Det[m_{11}]-1)^2\nonumber \\
&\geqslant& (\Det[K]-1)^2/4
\label{ineq1}\;,
\end{eqnarray}
where in the second identity the condition~(\ref{simpcond}) was
used.

As we shall see, with certain important exception one-mode
Gaussian channels~(\ref{linbos}) are unitarily equivalent to
transformations which admit physical representation with $\rho_b$
and $U_{ab}$ as in Eqs.~(\ref{sigmab}) and (\ref{couplingN}).

\subsection{Canonical form}\label{s:canonical}
Following Ref.~\cite{HOLEVOREP} any Gaussian
channel~(\ref{linbos1})  can be transformed (through unitarily
equivalence) into a simple canonical form. Namely, given a channel
$\Phi$ characterized by the vector $m$ and the matrices $K$,
$\alpha$ of Eq.~(\ref{linbos1}), one can find unitary operators
$U_a $ and $W_a$ such that the channel defined by the mapping
\begin{eqnarray}
\rho_a \longrightarrow \Phi^{(\text{can})} ( \rho_a) = W_a \;
\Phi( U_a \; \rho_a \; U_a^\dag) \; W_a^\dag \qquad \quad
\mbox{for all $\rho_a$,} \label{equivalent}
\end{eqnarray}
is of the form~(\ref{linbos1}) with $m=0$ and with $K$, $\alpha$
replaced, respectively, by the matrices  $K_{\text{can}}$,
$\alpha_{\text{can}}$ of Table~\ref{t:table}, i.e.
\begin{eqnarray}
\phi(\Phi^{(\text{can})}(\rho_a);z) =\phi(\rho_a; K_{\text{can}}
\cdot z)\; \exp[ -\frac{1}{2}\; z^T \cdot \alpha_{\text{can}}
\cdot z ]\;. \label{linbos2}
 \end{eqnarray}
An important consequence of Eq.~(\ref{linbos2}) is that to analyze
the weak-degradability properties of a one-mode Gaussian channel
it is sufficient to focus on the canonical map
$\Phi^{(\text{can})}$ which is unitarily equivalent to it (see
remark at the end of Sec.~\ref{s:uno}). Here we will not enter
into the details of the derivation of Eqs.~(\ref{equivalent}) and
(\ref{linbos2}), see Ref.~\cite{HOLEVOREP}.

\begin{table}[t]
\begin{tabular}  {cl|c|cc}
Channel $\Phi$ & &Class &  Canonical form  $\Phi^{(\text{can})}$ &  \\
$\mbox{Det} [K]$ & & &  $K_{\text{can}}$  & $\alpha_{\text{can}}$   \\
\hline \hline
$0$& $\mbox{rank} [K] = 0$ &  $A_1$  & $0$ &  $(N_0 +{1}/{2})\; \openone $  \\
$0$& $\mbox{rank} [K] = 1$& $A_2$ & $(\openone + \sigma_3) /2$ &  $(N_0 +{1}/{2})\; \openone $ \\
\hline $1$&$\mbox{rank}[\alpha] =1$ & $B_1$  & $\openone$ &
$(\openone - \sigma_3) /4$
\\
$1$ & $\mbox{rank}[\alpha] \neq 1$& $B_2$ & $\openone$ &  $N_0 \; \openone  $ \\
\hline
 $\kappa^2 \;\;\;(\kappa \neq 0,1) $ && $C$ &$ \kappa \;  \openone $
& $|\kappa^2-1| ( N_0 + 1/2) \; \openone$  \\
\hline
 $-\kappa^2 \;\; (\kappa \neq 0)$ && $D$ &$ \kappa \; \sigma_3 $
& $(\kappa^2+1) ( N_0 + 1/2) \; \openone$
\end{tabular}
\caption{Canonical form for one-mode Gaussian Bosonic channels.
In the first columns the properties of $K$ and $\alpha$  of the
map $\Phi$ are reported. In last two columns instead we give
the matrices $K_{\text{can}}$ and $\alpha_{\text{can}}$ of the canonical
form $\Phi^{(\text{can})}$
associated with $\Phi$ --- see Eqs.~(\ref{equivalent})
and (\ref{linbos2}).
 In these expressions $\sigma_3$ is the third Pauli matrix, $N_0$ is
a non-negative constant and $\kappa$ is a positive constant.
Notice that the constraint~(\ref{inequality}) is always satisfied.
In $B_1$ the free parameter $N_c$  has been set equal to $1/2$
--- see discussion below Eq.~(\ref{ALPHACAN}).
\label{t:table}} \end{table}

The dependence on the matrix $K_{\text{can}}$ of $\Phi^{(\text{can})}$
upon the parameters of $\Phi$ can be summarized as follows,
\begin{eqnarray}
K_{\text{can}} = \left\{ \begin{array}{cll}
\left\{ \begin{array}{lll}
\sqrt{\mbox{Det}[K]} \; \openone && \mbox{Det}[K]\geqslant 0 \\
\sqrt{|\mbox{Det}[K]|} \; \sigma_3 && \mbox{Det}[K]<0
\end{array} \right.
& &\mbox{rank}[K] \neq 1 \\ \\
(\openone + \sigma_3)/2 & &\mbox{rank}[K] = 1 \;,
\end{array}
\right.\label{KAPPACAN}
\end{eqnarray}
with $\sigma_3$ being the third Pauli matrix. Analogously for
$\alpha_{\text{can}}$ we have
\begin{eqnarray}
\alpha_{\text{can}} = \left\{ \begin{array}{cll}
\sqrt{\mbox{Det}[\alpha]} \; \openone & &\mbox{rank}[\alpha] \neq 1 \\ \\
 \; N_c \; (\openone - \sigma_3)/2 & &\mbox{rank}[\alpha] = 1 \;.
\end{array}
\right. \label{ALPHACAN}
\end{eqnarray}
The quantity $N_c$ is a free parameter which can set to any
positive value upon properly calibrating the unitaries $U_a$ and
$W_a$ of Eq.~(\ref{equivalent}). Following Ref.~\cite{HOLEVOREP}
we will assume $N_c=1/2$. Notice also that from
Eq.~(\ref{inequality}), $\mbox{rank}[\alpha]=1$ is only possible
for $\Det[K]=1$.

Equations~(\ref{KAPPACAN}) and (\ref{ALPHACAN}) show that only
the determinant and the rank of $K$ and $\alpha$ are relevant for defining
$K_{\text{can}}$ and $\alpha_{\text{can}}$.
Indeed one can verify that $K_{\text{can}}$ and $\alpha_{\text{can}}$
maintain the same determinant and rank of the original matrices $K$ and $\alpha$,
respectively. This is a consequence of the fact the $\Phi$ and
$\Phi^{(\text{can})}$ are connected through a  symplectic transformation for
which
$\mbox{Det}[K]$, $\mbox{Det}[\alpha]$,
$\mbox{rank}[K]$, and  $\mbox{rank}[\alpha]$
are invariant quantities.
[In particular $\mbox{Det}[ K]$ is directly
related with the invariant quantity $q$ analyzed in Ref.~\cite{CG}.]

\begin{figure}[t]
\centerline{\psfig{file=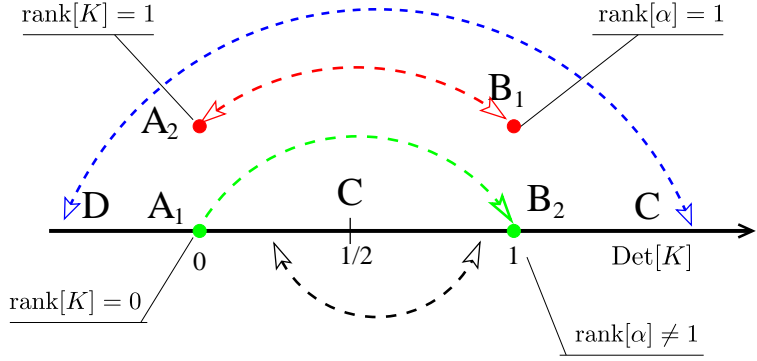,width= 10 cm}}
\caption{Pictorial representation of the classification in terms
of canonical forms of Table~\ref{t:table}. Depending on the values
of $\Det[K]$, $\rank[K]$ and $\rank[\alpha]$ any one-mode Gaussian
channel can be transformed to one of the channels of the scheme
through unitary transformations as in Eq.~(\ref{equivalent}). The
point on the thick oriented line for $\Det[K]<0$ represent the
maps of $D$, those with $\Det[K]>0$ and $\Det[K]\neq 1$ represent
$C$. The classes $A_{1,2}$ and $B_{1,2}$ are represented by the
four colored points of the graph. Notice that the channel $B_2$
and $A_1$ can be obtained as limiting cases of $D$ and $C$. The
dotted arrows connect channels which are weakly
complementary~(\ref{duedue})  of each others with respect to the
physical representations introduced in Sec.~\ref{sec:single}. For
instance the weakly complementary of  $B_1$ is channel of the
class $A_2$ (and vice-versa) --- see Sec.~\ref{s:weakc} and
Table~\ref{t:table10} for details. Notice that the weakly
complementary channel of $A_1$ belongs to $B_2$. However, not all
the channels of $B_2$ have weakly complementary channels which are
in $A_1$ --- see Sec.~\ref{sec:b2}.} \label{f:scheme}
\end{figure}

The six inequivalent canonical forms of Table~\ref{t:table} follow
by parametrizing the value of $\sqrt{\mbox{Det}[\alpha]}$ to
account for the constraints imposed by  the
inequality~(\ref{inequality}). It should be noticed that to
determine which class a certain channel belongs to, it is only
necessary to know if $\Det[K]$ is null, equal to $1$, negative or
positive ($\neq 1$). If $\Det[K]=0$ the class is determined by the
rank of the matrix. If $\Det[K]=1$ the class is determined by the
rank of $\alpha$ (see Fig.~\ref{f:scheme}). Within the various
classes, the specific expression of the canonical form depends
then upon the effective values of $\Det[K]$ and $\Det[\alpha]$. We
observe also that the class  $A_1$ can be obtained as a limiting
case (for $\kappa\rightarrow 0$) of the maps of class $C$ or $D$.
Analogously the class $B_2$ can be obtained as a limiting case of
the maps of class $C$.  Indeed consider the channel with
$K_{\text{can}} = \kappa \openone$ and $\alpha_{\text{can}} 
=|\kappa^2 -1| (N_0^\prime +1/2) \openone$ 
with $N_0^\prime={N_0}/({|\kappa^2 -1|}) -1/2$, with $N_0$ and $\kappa$ positive
($\kappa \neq 0,1$). For $\kappa$
sufficiently close to $1$, $N_0^\prime$ is positive and the maps
belongs to the class $C$ of Table~\ref{t:table}. Moreover in the
limit of $\kappa\rightarrow 1$ this channel yields the map $B_2$.

Finally it is interesting to study how the canonical forms of
Table~\ref{t:table} compose under the product~(\ref{composition}).
A simple calculation shows that the following rules apply
\begin{eqnarray}
\label{compo1}
\begin{array}{c|cccccc}
 \circ     & A_1 & A_2 & B_1 & B_2 & C & D \\ \hline
A_1 & A_1 & A_1 & A_1 & A_1 & A_1 & A_1 \\
A_2 & A_1 & A_2 & A_2 & A_2 & A_2 & A_2 \\
B_1 & A_1 & A_2 & B_1 & B_{1}/B_{2} & C & D \\
B_2 & A_1 & A_2 & B_{1}/B_{2} & B_{2} & C & D \\
C & A_1 & A_2 & C & C & B_2/C & D \\
D & A_1 & A_2 & D & D & D & C
\end{array}
\end{eqnarray}
In this table, for instance, the element on the row 2 and column 3
represents class (i.e. $A_2$) associated to the product $\Phi^{\prime\prime}\circ \Phi^{\prime}$
between a channel $\Phi^{\prime}$ of $B_1$ and a channel $\Phi^{\prime\prime}$
of $A_2$.
Notice that the canonical form of the products $B_1\circ B_2$, $B_2\circ B_1$ and $C\circ C$
is not uniquely defined.
In the first case in fact, even though the determinant of the matrix $K$
of Eq.~(\ref{composition}) is one, the rank of the corresponding $\alpha$ might be one or
different from one depending on the parameters of the two ``factor'' channels: consequently
the  $B_1\circ B_2$ and $B_2\circ B_1$ might belong either to $B_1$ or to $B_2$.
In the case of $C\circ C$ instead it is possible that the resulting channel will have
$\Det[K]=1$ making it a $B_2$ map. Typically however $C\circ C$ will be a map of $C$.
Composition rules analogous to those reported here have been extensively analyzed
in Refs.~\cite{CG,ENTROPY,GL}.

\subsection{Single-mode physical representation of the canonical
forms}
\label{sec:single}

Apart from the case $B_2$ that will be treated separately (see
next section), all canonical transformations of
Table~\ref{t:table} can be expressed as in Eq.~(\ref{output}),
i.e. through a physical representation~(\ref{HGCuno}) with
$\rho_b$ being a thermal state~(\ref{sigmab}) of a single external
Bosonic mode $B$ and $U_{ab}$ being a linear
transformation~(\ref{couplingN})\footnote{The exceptional role of $B_2$
 corresponds to the fact that any one-mode Bosonic
Gaussian channel can be represented as a unitary coupling with
a single-mode environment
plus an additive classical noise (see next section and Ref.~\cite{REV}).}.
To show this it is sufficient to
verify that, for each of the classes of Table~\ref{t:table} but
$B_2$, there exists  a non-negative number $N$ and a symplectic matrix
$M$  such that Eq.~(\ref{output}) gives the
mapping~(\ref{linbos2}). This yields the conditions
\begin{eqnarray}
m_{11} &=& K_{\text{can}}\;, \label{m11}\\
m_{12}  &=& O \;\sqrt{ \frac{\alpha_{\text{can}}}{N+1/2}}\;,
\label{m12}
\end{eqnarray}
with $O^T=O^{-1}$ being an orthogonal $2\times 2$ matrix to be
determined through the symplectic condition
\begin{eqnarray}
\Det[m_{11}] + \Det[m_{12}] =1
\label{simpcond} \;,
\end{eqnarray}
 which guarantees
that $U_{ab}^\dag Q_a U_{ab}$ and $U_{ab}^\dag P_a U_{ab}$ satisfy
canonical commutation relations. It is worth noticing that once
$m_{11}$ and $m_{12}$ are determined within the
constraint~(\ref{simpcond}) the remaining blocks (i.e. $m_{21}$
and $m_{22}$) can always be found in order to satisfy the
remaining symplectic conditions of  $M$. An explicit example will
be  provided in few paragraphs. For the classes $A_1$, $A_2$,
$B_1$, $D$, and $C$ with $\kappa<1$, Eqs.~(\ref{m12}) and
(\ref{simpcond}) can be solved by choosing $O = \openone$ and
$N=N_0$. Indeed for $B_1$ the latter setting is not necessary. Any
non-negative number will do the job: thus we choose  $N=0$ making the
density matrix $\rho_b$ of Eq.~(\ref{sigmab}) the vacuum of the
$B$. For $C$ with $\kappa>1$ instead a solution is obtained by
choosing $O = \sigma_3$ and  again $N=N_0$. The corresponding
transformations~(\ref{couplingN})  for $Q_a$ and $P_a$
(together with the choice for $N$) are  summarized below.
\begin{eqnarray}
\begin{array}{cc|c|ccc}
\mbox{Class} & & \rho_b &U_{ab}^\dag \; Q_a
\; U_{ab} && U_{ab}^\dag \; P_a \; U_{ab}  \\
\hline
A_1 && \mbox{thermal} (N=N_0) & Q_b && P_b \\
A_2 && \mbox{thermal} (N=N_0) & Q_a + Q_b && P_b \\
B_1 && \mbox{vacuum} (N=0)& Q_a && P_a+ P_b   \\
C   & \kappa<1 & \mbox{thermal} (N=N_0)&\kappa \; Q_a +
\sqrt{1-\kappa^2} \; Q_b
&& \kappa \; P_a + \sqrt{1-\kappa^2} \; P_b  \\
C  & \kappa> 1 & \mbox{thermal} (N=N_0) &\kappa \; Q_a +
\sqrt{\kappa^2-1} \; Q_b &&
\kappa \; P_a - \sqrt{\kappa^2-1} \; P_b\\
D &  &  \mbox{thermal} (N=N_0) & {\kappa} \; Q_a + \sqrt{\kappa^2
+1} \; Q_b &&
 - {\kappa} \; P_a + \sqrt{\kappa^2+1} \; P_b \;.
\nonumber
\end{array}
\end{eqnarray}
To complete the definition of the unitary operators $U_{ab}$ we
need to provide also the transformations of $Q_b$ and $P_b$. This
corresponds to fixing the blocks $m_{21}$ and $m_{22}$ of $M$ and
cannot be done uniquely: one possible choice is presented in the
following table
\begin{eqnarray}
\begin{array}{cc|ccc}
\mbox{Class} & &U_{ab}^\dag \; Q_b
\; U_{ab} &&U_{ab}^\dag \; P_b \; U_{ab}  \\
\hline
A_1 &&  Q_a && P_a \\
A_2 &&  Q_a && P_a - P_b \\
B_1 &&  Q_a - Q_b && -P_b   \\
C   & \kappa<1 &  \sqrt{1-\kappa^2} \; Q_a- \kappa \; Q_b
&&  \sqrt{1-\kappa^2} \; P_a- \kappa \; P_b   \\
C  & \kappa> 1 &   \sqrt{\kappa^2-1} \; Q_a+  \kappa \; Q_b  &&
 - \sqrt{\kappa^2-1} \; P_a+ \kappa \; P_b \\
D &  &   \sqrt{\kappa^2 +1} \; Q_a +  {\kappa} \; Q_b &&
  \; \sqrt{\kappa^2+1} \; P_a  - {\kappa} \; P_b \;.
\end{array}
\nonumber
\end{eqnarray}
The above definitions  make explicit the fact that the canonical
form $C$ represents attenuator ($\kappa<1$)  and amplifier
($\kappa>1$)  channel~\cite{HW}. We will see in the following
sections that the class $D$ is formed by the weakly complementary
of the amplifier channels of the class $C$. For the sake of
clarity the explicit expression for the matrices $M$ of the
various classes has been reported in App.~\ref{appendiceM}.

Finally it is important to notice that the above physical
representations are equivalent to  Stinespring representations
only when the average photon number $N$ of $\rho_b$ nullifies. In
this case the environment $B$ is represented by a pure input state
(i.e. the vacuum). According to our definitions this is always the
case for the canonical form
 $B_1$ while  for the canonical forms $A_1$, $A_2$, $C$ and $D$
it happens for $N_0=0$.

\subsection{The class $B_2$: additive classical noise
channel}\label{sec:b2}

As mentioned in the previous section the class $B_2$ of
Table~\ref{t:table} must be treated separately. The map $B_2$
corresponds\footnote{This can be
seen  for instance by evaluating the characteristic function of
the state~(\ref{additive}) and comparing it with
Eq.~(\ref{linbos2}).}
 to the additive classical noise channel~\cite{HW} defined by
\begin{eqnarray}
\Phi(\rho_a) = \int d^2 z \; p(z) \; V_a(z) \; \rho_a \; V_a(-z)
\label{additive}\;,
\end{eqnarray}
with $p(z) = (2 \pi N_0)^{-1} \; \exp[-|z|^2/(2N_0) ]$
which, in Heisenberg picture, can be seen as a random
shift of the annihilation operator $a$.

These channels
 admit a natural physical representation which
involve two environmental modes in a pure state (see
Ref.~\cite{HOLEVOREP} for details) but do not have a physical
representations~(\ref{HGCuno}) involving a single environmental
mode. This can be verified by noticing that in this case, from
Eqs.~(\ref{m11}) and (\ref{m12}) we get
\begin{eqnarray}
m_{11} &=& \openone \label{m11B2} \\
m_{12} &=& \sqrt{N_0/(N+1/2)} \; O\;, \label{m12B2}
\end{eqnarray}
which  yields
\begin{eqnarray}
\Det[m_{11}] + \Det[m_{12}] = 1 \pm N_0/(N+1/2) \;,
\end{eqnarray}
independently of the choice of the orthogonal matrix
$O$\footnote{This follows from the fact that $\Det[O]=\pm 1$ since
$O^T =O^{-1}$.}. Therefore, apart from the trivial case $N_0=0$,
the only solution to the constraint~(\ref{simpcond}) is by taking
the limit $N\rightarrow \infty$. This would correspond to
representing the channel $B_2$ in terms of a linear coupling with
a single-mode thermal state $\rho_b$ of ``infinite'' temperature.
Unfortunately this is not a well defined object. However we can
use the ``asymptotic'' representation described at the end of
Sec.~\ref{s:canonical} where it was shown how to obtain $B_2$ as
limiting case of $C$ class maps, to claim at least that there
exists a one-parameter family of one-mode Gaussian channels which
admits single-mode physical representation and which converges to
$B_2$.

\begin{table}[t!]
\begin{tabular}  {cc|cc|c}
Class of ${\Phi}$  & & Weak complementary channel $\tilde{\Phi}$ &
 & Class of $\tilde{\Phi}$ \\
&& $K$ &$\alpha$ & \\
\hline \hline $A_1$ &&  $\openone$ & $0$ & $B_2$ \\ \hline $A_2$
&&  $\openone$ & $(N_0+1/2) \; (\openone -\sigma_3)/2$ & $B_1$ \\
\hline $B_1$ &&  $(\openone+\sigma_3)/2$ & $\openone/2$ & $A_2$  \\
\hline $C$   & $\kappa<1$ & $ \sqrt{1-\kappa^2} \; \openone$
& $ k^2 (N_0+1/2) \openone $ & $C\;\; (\kappa<1)$ \\
$C$  & $\kappa> 1$ &  $ \sqrt{\kappa^2-1} \; \sigma_3$ & $\kappa^2
(N_0+1/2) \; \openone $ & $D$\\ \hline $D$ &  &  $ \sqrt{\kappa^2
+1} \; \openone$ & $  \kappa^2 (N_0+1/2)\; \openone $ & $C \;\;
(\kappa>1)$
\end{tabular}
\caption{Description of the weakly complementary~(\ref{duedue}) of
the canonical forms $A_1$, $A_2$, $B_1$, $C$ and $D$ of
Table~\ref{t:table} constructed from the physical
representations~(\ref{HGCuno}) given in Sec.~\ref{sec:single}. In
the first column is indicated the class of $\Phi$. In the central
columns instead is given a description of $\tilde{\Phi}$ in terms
of the representation~(\ref{linbos1}). Finally in the last column
is reported the canonical form corresponding to the map
$\tilde{\Phi}$. In all cases the identification is immediate: for
instance the canonical form of the map $\tilde{\Phi}_{A_1}$
belongs to the class $B_2$, while the canonical form of the map
$\tilde{\Phi}_{D}$ is the class $C$ with $\Det[K_{\text{can}}]>1$.
In the case of $\tilde{\Phi}_{A_2}$ the identification with the
class $B_1$ was done by exploiting the possibility freely varying
$N_c$ of Eq.~(\ref{ALPHACAN})
--- see Ref.~\cite{HOLEVOREP}.
A pictorial representation of the
above weak-degradability connections is given in Fig.~\ref{f:scheme}.
\label{t:table10}}
\end{table}

\section{Weak-degradability of one-mode Gaussian channels}\label{s:DEGRADABILE}

In the previous section we have seen that all one-mode Gaussian
channels are unitarily equivalent to one of the canonical forms of
Table~\ref{t:table}. Moreover we verified that, with the exception
of the class $B_2$, all the canonical forms admits a physical
representation~(\ref{HGCuno}) with $\rho_b$ being a thermal state
of a single environmental mode and $U_{ab}$ being a linear
coupling. Here we will use such representations to construct the
weakly complementary~(\ref{duedue}) of these channels and to study
their weak-degradability  properties.

\subsection{Weakly complementary channels}\label{s:weakc}
In this section  we construct the weakly complementary channels
$\tilde{\Phi}$ of the class $A_1$, $A_2$, $B_1$, $C$ and $D$
starting from their single-mode physical
representations~(\ref{HGCuno}) of Sec.~\ref{sec:single}. Because
of the linearity of $U_{ab}$ and the fact that $\rho_b$ is
Gaussian, the channels $\tilde{\Phi}$ are Gaussian. This can be
seen for instance by computing the characteristic
function~(\ref{characte}) of the output state
$\tilde{\Phi}(\rho_a)$
\begin{eqnarray}
\phi(\tilde{\Phi}(\rho_a); z) &=& \Tr_b[ \tilde{\Phi}(\rho_a) \;
V_b(z) ] = \Tr_{b} [ \rho_a \; \tilde{\Phi}_H(V_b(z)) ]
\nonumber \\
&=& \phi( \rho_a; m_{21} \cdot z) \; \exp[ - \frac{1}{2} ( N+ 1/2)
\; | m_{22} \cdot z|^2 ] \label{charN} \;,
\end{eqnarray}
where $m_{21}$, $m_{22}$ are the blocks elements of the
matrix $M$ of Eq.~(\ref{matricem}) associated with the transformations
$U_{ab}$, and with $N$ being the average photon number of $\rho_b$
(the values of these quantities are given in the tables of
Sec.~\ref{sec:single} --- see also App.~\ref{appendiceM}). By
setting $m=0$, $K= m_{21}$ and $\alpha= (N+1/2) \; m_{22}^T \;
m_{22}$, Eq.~(\ref{charN}) has the same structure~(\ref{linbos1})
of the one-mode Gaussian channel  of $A$. Therefore by cascading
$\tilde{\Phi}$ with an isometry which exchanges $A$ with $B$ (see
Refs.~\cite{SARO,CG}) we can then treat $\tilde{\Phi}$ as an
one-mode Gaussian channel operating on $A$ (this is possible
because both $A$ and $B$ are Bosonic one-mode systems). With the
help of Table~\ref{t:table} we can then determine which classes
can be associated with the transformation~(\ref{charN}). This is
summarized in Table~\ref{t:table10}.

\subsection{Weak-degradability properties}\label{sec:wdp}

\begin{figure}[t]
\centerline{\psfig{file=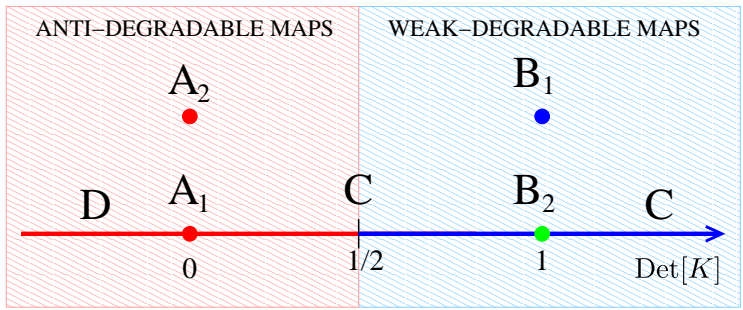,width= 10 cm}}
\caption{Pictorial representation of the weak-degradability
regions for one-mode Gaussian channels. All canonical forms with
$\Det[K]\leqslant 1/2$ are anti-degradable: this includes the
classes $A_{1}$, $A_2$, $D$ and part of the $C$. The remaining
(with the exception of $B_2$) are instead weakly degradable.
Moreover $B_1$ is also degradable  in the sense of
Ref.~\cite{DEVSHOR}. The same holds for channels of canonical form
$C$ with $N_0=0$: the exact expression for the quantum capacity of
these channels has been given in Ref.~\cite{WOLF}.}
\label{f:scheme2}
\end{figure}

Using the compositions rules of Eqs.~(\ref{composition}) and
(\ref{compo1}) it is easy to verify that the canonical forms
$A_1$, $A_2$, $D$ and $C$ with  $\kappa\leqslant \sqrt{1/2}$ are
anti-degradable~(\ref{wwwanti}). Vice-versa one can  verify that
the canonical forms $B_1$ and $C$ with $\kappa\geqslant
\sqrt{1/2}$ are weakly degradable~(\ref{wwwweak})
--- for $C$, $D$ and $A_1$ these results have been proved in Ref.~\cite{CG}.
Through unitary equivalence this can be summarized by saying that
all one-mode Gaussian channels~(\ref{linbos1}) having
$\Det[K]\leqslant 1/2$ are anti-degradable, while the others (with
the exception of the channels belonging to $B_2$) are weakly
degradable (see Fig.~\ref{f:scheme2}).

In the following we verify the above relations by explicitly
constructing the connecting channels $\Psi$ and $\overline{\Psi}$
of Eqs.~(\ref{wwwweak}) and (\ref{wwwanti}) for each of the
mentioned canonical forms. Indeed one has:

\begin{itemize}
\item For a channel  $\Phi$ of standard form $A_1$ or $A_2$, anti-degradability
can be shown by simply taking $\overline{\Psi}$ of
Eq.~(\ref{wwwanti}) coincident with the channel $\Phi$. The result
immediately follows from the composition rule~(\ref{composition}).
\item For a channel $\Phi$ of $B_1$, weak-degradability comes by assuming the map $\Psi$
to be equal to the weakly complementary channel $\tilde{\Phi}$ of
$\Phi$ (see Table~\ref{t:table10}). As pointed out in
Ref.~\cite{HOLEVOREP} this also implies the degradability of
$\Phi$ in the sense of Ref.~\cite{DEVSHOR}. Let us remind that for
$B_1$ the physical representation given in Sec.~\ref{sec:single}
was constructed with an environmental state $\rho_b$ initially
prepared in the vacuum state, which is pure. Therefore in this
case our representation gives rise to a Stinespring dilation.
\item For a channel $\Phi$ of the class $C$ with $K_{\text{can}}= \kappa \; \openone$
and $\alpha_{\text{can}} =|\kappa^2 -1| (N_0+1/2) \openone$  we have the following three possibilities:
\begin{itemize}
\item If $\kappa \leqslant \sqrt{1/2}$ the channel is anti-degradable and the connecting map $\overline{\Psi}$
is a channel of $C$ characterized by $K_{\text{can}}= \kappa^\prime \; \openone$
and $\alpha_{\text{can}} =(1-(\kappa^\prime)^2)  (N_0+1/2) \openone$ with
$\kappa^\prime = \kappa/\sqrt{1-\kappa^2} <1$.
\item If $\kappa \in [\sqrt{1/2},1[$ the channel is weakly degradable and the connecting map ${\Psi}$
is again a channel of $C$ defined as in the previous case but with $\kappa^\prime
= \sqrt{1-\kappa^2}/\kappa <1$. For $N_0=0$ the channel is also
degradable~\cite{DEVSHOR} since our physical
representation is equivalent to a Stinespring representation.

\item If $\kappa >1$ the channel is weakly degradable and the connecting map ${\Psi}$
is a channel of $D$ with $K_{\text{can}}= \kappa^\prime \; \openone$
and $\alpha_{\text{can}} =((\kappa^\prime)^2-1)  (N_0+1/2) \openone$
with $\kappa^\prime =\sqrt{k^2-1}/k$. As in the previous case,
 for $N_0=0$ the channel is also
degradable~\cite{DEVSHOR}.

\end{itemize}
\item For a channel $\Phi$ of $D$ with $K_{\text{can}} = \kappa \; \sigma_3$ and $\alpha_{\text{can}}
= (\kappa^2 +1) (N_0+1/2) \openone$ ($\kappa>0$ and
$N_0\geqslant0$) we can prove anti-degradability by choosing
$\overline{\Psi}$ of Eq.~(\ref{wwwanti}) to be yet another maps of
$D$ with $K_{\text{can}} =\kappa^\prime  \; \sigma_3$ and
$\alpha_{\text{can}} = ((\kappa^\prime)^2 +1) (N_0+1/2) \openone$
where $\kappa^\prime=\kappa/\sqrt{\kappa^2 +1}$. From
Eq.~(\ref{composition}) and Table~\ref{t:table10}  it then follows
that $\Psi \circ \tilde{\Phi}$ is indeed equal to $\Phi$.
\end{itemize}

Concerning the case $B_2$ it was shown in  Ref.~\cite{HOLEVOREP}
that the channel is neither anti-degradable nor degradable in the
sense of~\cite{DEVSHOR} (apart from the trivial case $N_0=0$ which
corresponds to the identity map). On the other hand one can use
the continuity argument given in Sec.~\ref{sec:b2} to claim that
the channel $B_2$ can be arbitrarily approximated with maps which
are weakly degradable (those belonging to $C$ for instance).

\section{One-mode Gaussian channels with $\Det[K]>1/2$ and having
null quantum capacity}\label{sec:null}

In the previous section we saw that all channels~(\ref{linbos1})
with $\Det[K]\leqslant 1/2$ are anti-degradable. Consequently
these channel must have null quantum capacity~\cite{CG,SARO}. Here
we go a little further showing that  the set of the maps
(\ref{linbos1}) which can be proved to have null quantum capacity
include also some maps with $\Det[K] >1/2$. To do this we will use
the following simple fact:

{\em Let be $\Phi_1$ a quantum channel with null quantum capacity
and let be $\Phi_2$ some quantum channel. Then the composite
channels $\Phi_1\circ \Phi_2$ and $\Phi_2 \circ \Phi_1$ have null
quantum capacity.}

The proof of this property follows by interpreting $\Phi_2$ as a
quantum operation performed either at the decoding or at encoding
stage of the channel $\Phi_1$. This shows that the quantum
capacities of $\Phi_1\circ \Phi_2$ and $\Phi_2 \circ \Phi_1$
cannot be greater than the capacity of $\Phi_1$ (which is null). In the following
we will present two cases where the above property turns out to
provide some nontrivial results.

\subsection{Composition of two class $D$ channels}\label{sec:2D}

\begin{figure}[t]
\centerline{\psfig{file=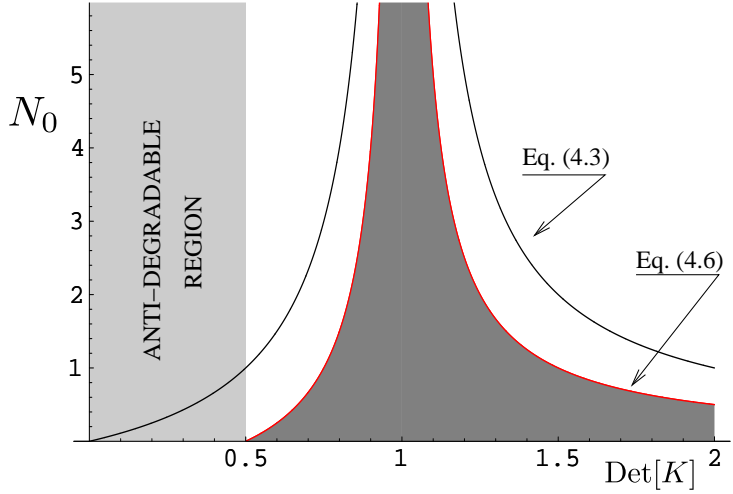,width= 10 cm}}
\caption{The dark-grey area of the plot is the region of the parameters $N_0$ and $\Det[K]=\kappa^2$
where a channel with canonical form $C$ can have not null quantum capacity.
For $\Det[K]<1/2$ the channel is anti-degradable.  In the remaining white area
the quantum capacity is null since these maps can be  obtained by a composition
of channels one of which being anti-degradable.
The curve in black refers to the bound of Eq.~(\ref{enneN}).
 The contour of the dark-grey area is
instead given by Eq.~(\ref{enneNN}).}
\label{f:plot}
\end{figure}

We observe that according to composition rule~(\ref{compo1})
the combination of any
two channels $\Phi_1$ and $\Phi_2$ of $D$ produces a map $\Phi_{21} \equiv \Phi_2\circ \Phi_1$
 which is in the class $C$.
Since the class $D$ is anti-degradable the resulting channel
must have null quantum capacity.
Let then $\kappa_j \sigma_3$ and $(\kappa^2_j +1) (N_j+1/2)
\openone$ be the matrices $K_{\text{can}}$ and
$\alpha_{\text{can}}$ of the channels $\Phi_j$, for $j=1,2$. From
Eq.~(\ref{composition}) one can then verify that $\Phi_{21}$ has
the canonical form $C$ with parameters
\begin{eqnarray}
\kappa &=& \kappa_1 \kappa_2 \;,\label{kappa} \\
N_0  &=& \frac{(\kappa_2^2+1) N_2 + \kappa_2^2 (\kappa_1^2+1) N_1 }{|\kappa_1^2\kappa_2^2 -1|}
+ \frac{1}{2} \left( \frac{\kappa^2_1\kappa_2^2 + 2 \kappa_2^2 +1}{ |\kappa_1^2\kappa_2^2 -1|} -1
\right) \;. \label{enne}
\end{eqnarray}
Equation~(\ref{kappa}) shows that by varying $\kappa_j$, $\kappa$
can take any positive values: in particular it can be greater than
$\sqrt{1/2}$ transforming  $\Phi_{21}$ into a
channel which does not belong to the anti-degradable area of
Fig.~\ref{f:scheme2}. On the other hand, by varying the $N_j$ and
$\kappa_2$, but keeping the product $\kappa_1\kappa_2$ fixed, the
parameter $N_0$ can assume any value satisfying the inequality
\begin{eqnarray}
N_{0}  &\geqslant&
\frac{1}{2} \left( \frac{\kappa^2 +1}{ |\kappa^2 -1|} -1
\right) \;. \label{enneN}
\end{eqnarray}
We can therefore conclude that all channels $C$ with
$\kappa$ and $N_0$ as in Eq.~(\ref{enneN}) have null quantum capacity
--- see Fig.~\ref{f:plot}.
A similar bound was found in a completely different way  in Ref.~\cite{HW}.

\subsection{Composition of two class $C$ channels}\label{sec:2C}

Consider now the composition
of two class $C$ channels, i.e. $\Phi_{1}$ and $\Phi_2$, with
one of them (say $\Phi_2$) being anti-degradable.

Here, the canonical form of $\Phi_1$ and $\Phi_2$ have matrices
 $K_{\text{can}}$ and $\alpha_{\text{can}}$
given by $K_i=\kappa_j \openone$ and $\alpha_j= |\kappa^2_j -1| (N_j+1/2) \openone$,
where for $j=1,2$,  $N_j$ and $\kappa_j$ are positive numbers, with
$\kappa_{1}\neq 0,1$ and with $\kappa_2 \in ]0,\sqrt{1/2}]$ (to ensure anti-degradability).
From Eq.~(\ref{composition}) follows then that
the composite map $\Phi_{21}= \Phi_2 \circ \Phi_1$ has still a $C$ canonical form
with parameters
\begin{eqnarray}
\kappa &=& \kappa_1 \kappa_2 \;,\label{kappa1} \\
N_0  &=& \frac{|\kappa_2^2-1| N_2 + \kappa_2^2 |\kappa_1^2-1| N_1 }{|\kappa_1^2\kappa_2^2 -1|}
+ \frac{1}{2} \left( \frac{\kappa^2_2|\kappa_1^2-1| + | \kappa_2^2 -1|}{ |\kappa_1^2\kappa_2^2 -1|} -1
\right) \;. \label{enne111}
\end{eqnarray}
As in the previous example, $\kappa$ can assume any positive value.
Vice-versa keeping $\kappa$ fixed, and varying $\kappa_1 >1$ and  $N_{1,2}$
it follows that $N_0$ can take any values which satisfy the inequality
  \begin{eqnarray}
N_{0}  &\geqslant&
\frac{1}{2} \left( \frac{\kappa^2}{ |\kappa^2 -1|} -1
\right) \;. \label{enneNN}
\end{eqnarray}
We can then conclude that all maps $C$ with $\kappa$ and $N_0$ as
above must possess null quantum capacity. The result has been
plotted in Fig.~\ref{f:plot}. Notice that the
constraint~(\ref{enneNN}) is an improvement with respect to the
constraint of Eq.~(\ref{enneN}).

\section{Conclusion}\label{sec:con}

In this paper we provide a full weak-degradability classification
of one-mode Gaussian channels by exploiting the canonical form
decomposition of Ref.~\cite{HOLEVOREP}. Within this context we
identify those channels which are anti-degradable. By exploiting
composition rules of Gaussian maps, this allows us to strengthen
the bound for one-mode Gaussian channels which have not null
quantum capacity.

F.C. and V.G. thank the Quantum Information research program of
Centro di Ricerca Matematica Ennio De Giorgi of Scuola Normale
Superiore for financial support. A. H. acknowledges hospitality of
Centre for Quantum Computation, Department of Applied Mathematics
and Theoretical Physics, Cambridge University.

\appendix

\section{The matrix $M$}\label{appendiceM}

Here we give the explicit expressions of the matrix $M$ of
Eq.~(\ref{matricem}) associated with the physical representations
of the classes $A_1$, $A_2$, $B_1$, $C$ and $D$, discussed in
Sec.~\ref{sec:single}. They are,
\begin{eqnarray*}
M_{A_1} \equiv \left( \begin{array}{cc|cc}
0&0& 1&0 \\
0&0& 0&1 \\ \hline
1&0& 0&0 \\
0&1& 0&0 \\
\end{array} \right)\;,
\quad
M_{A_2} \equiv \left( \begin{array}{cc|cr}
1&0& 1&\!\!0 \\
0&0& 0&\!\!1 \\ \hline
1&0& 0&\!\!0 \\
0&1& 0&\!\!-1 \\
\end{array} \right) \;,
\quad
M_{B_1} \equiv \left( \begin{array}{cc|rr}
1&0& \!\!1&\!\!0 \\
0&1& \!\!0&\!\!0 \\ \hline
0&0& \!\!-1&\!\!0 \\
0&1& \!\!0&\!\!-1 \\
\end{array} \right) \;,
\end{eqnarray*}
\begin{eqnarray*}
M_{C} \equiv \left( \begin{array}{cc|cc}
k&0& \sqrt{1-k^2}& 0 \\
0&k& 0& \sqrt{1-k^2} \\ \hline
\sqrt{1-k^2}&0& -k& 0 \\
0&\sqrt{1-k^2}&  0& -k \\
\end{array} \right) \qquad \mbox{(for $\kappa<1$),}
\end{eqnarray*}
\begin{eqnarray*}
M_{C} \equiv \left( \begin{array}{cc|cc}
k&0& \sqrt{k^2-1}& 0 \\
0&k& 0& - \sqrt{k^2-1} \\ \hline
\sqrt{k^2-1}&0& k&0 \\
0&-\sqrt{k^2-1}& 0&k \\
\end{array} \right) \qquad \mbox{(for $\kappa>1$),}
\end{eqnarray*}
\begin{eqnarray*}
M_{D} \equiv \left( \begin{array}{cc|cc}
k&0& \sqrt{k^2+1}& 0 \\
0&-k& 0&  \sqrt{k^2+1} \\ \hline
\sqrt{k^2+1}&0& k&0 \\
0&\sqrt{k^2+1}& 0&-k \\
\end{array} \right) \;.
\end{eqnarray*}

\end{document}